\documentclass[10pt]{iopart}
\usepackage{amsfonts}
\usepackage{amsthm}
\usepackage{amssymb}
\usepackage{graphicx}
\newcommand{\be}{\begin{equation}}
\newcommand{\ee}{\end{equation}}
\def\beqa{\begin{eqnarray}}
\def\eeqa{\end{eqnarray}}
\def\beq{\begin{equation}}
\def\eeq{\end{equation}}

\def\cqg{{\it Class. Quantum Grav.}\ }

\def\grg{{\it Gen. Relativ. Grav.}\ }

\let\gam=w

\renewcommand{\epsilon}{\varepsilon}

\def\rf#1{(\ref{#1})}
\begin{document}
\title{Cosmological dynamics of exponential gravity}
\author{M. Abdelwahab\S, S Carloni\S, P K. S. Dunsby\S\dag }

\address{\S\ Department of Mathematics and Applied\
Mathematics, University of Cape Town, South Africa.}

\address{\dag\ South African Astronomical Observatory,
Observatory Cape Town, South Africa.}

\begin{abstract}
We present a detailed investigation of the cosmological  dynamics based on 
$\exp\, (-R/{\Lambda})$ gravity. We apply the dynamical system approach to both the 
vacuum and matter cases and obtain exact solutions and their stability in the finite and 
asymptotic regimes. The results show that cosmic histories exist which  admit a double
de-Sitter phase which could be useful for describing the early and
the late-time accelerating universe.
\end{abstract}
\section{Introduction}
In recent years several observational surveys appear  to show that
the universe is currently undergoing an  accelerated expansion phase
(for a review see \cite{b}). On the other hand, astrophysical
observations on galactic scales gave us a clear indication that the
amount of the luminous matter is not enough to account for the
rotation curves of galaxies. These remarkable observations have
radically changed  our ideas about the evolution of the universe. In
fact, the only way in which this behaviour can be explained within
the standard Friedmann general relativistic cosmology framework is
to invoke  two dark matter components, one with a conventional dust
equation of state ($w=0$) needed to fit astrophysical data on
galactic scales (Dark Matter) and the other with a more exotic
equation of state $(w < -\frac{1}{3})$ needed to explain the current
accelerated expansion phase of the universe (Dark Energy).  There is
little doubt that  unraveling the nature of Dark matter and Dark
Energy is currently one of the most important problems in
theoretical physics.

One approach to the problem of Dark Energy that has received
considerable attention in the last few years is the modification of
General Relativity on cosmological scales.  In this approach one
supposes that Dark energy is a manifestation of a non-Einsteinian
behaviour of the gravitational interaction rather than a new form of
energy density. The introduction of corrections to the Hilbert
Einstein action and their effects have been studied for long time
and are believed to be unavoidable when the quantum nature of the
universe is introduced in General Relativity (GR).

In the last few years many different extended versions of the
Einstein theory of gravity have been proposed. One of the most
studied approaches is higher order theories of gravity
$[13-42]$, in which the gravitational action is nonlinear in the Ricci
curvature and/or it's derivatives \cite{c,d,e}. These theories have a 
number of interesting features on cosmological and
astrophysical scales. In fact they are known to admit natural
inflation phases \cite{Starobinsky} and to explain the flattening of
the galactic rotation curves \cite{f}. Another very interesting
feature of these models is that the higher order corrections to
Hilbert-Einstein action can be viewed as an effective fluid which
can mimic the properties of Dark Energy \cite{Capozziello}.

One of the the main problems that occurs in the study of higher
order theories of gravity is that finding exact cosmological solutions is 
extremely difficult due to the high degree of non-linearity
exhibited by these theories. This problem can be partially addressed
using a suitable choice of generalized coordinates in which the
field equations can be written as a system of first-order autonomous
differential equations together with a constraint equation \cite{g}.
In this way we can exploit the  methods of dynamical systems theory
\cite{a} in order to both understand the qualitative behavior of the
cosmological dynamics and obtain special exact solutions of the
cosmological equations.  The general approach allowing one to 
analyze higher order gravity with dynamical systems techniques 
has been presented elsewhere \cite{generaldynsys}.

In this paper we will apply this approach to an important class of
theories with Lagrangian density
\begin{eqnarray}
\mathcal{A}=\int d^{4}x {\cal L} = \int d^{4}x \sqrt{-g} \left [ e^{-\frac{R}{\Lambda}}+L_{M}\right]\;,
\end{eqnarray}
where $\Lambda$ is the cosmological constant. This Lagrangian has
some interesting  features, first of all its treatment is equivalent to one 
involving a combination of powers of the  Ricci scalar and, secondly
it reduces to
\begin{equation}
\exp\left(-\frac{R}{\Lambda}\right)=  1 - \frac{R}{\Lambda}+ O[R^{2}] \;,
\end{equation}
In the small curvature limit, which is equivalent to Hilbert-Einstein action.

This paper has been arranged as follows. In section $2$, we present
the basic equations of the model. In section $3$ and $4$ we find
exact solutions and their stability in the vacuum and matter  cases
respectively. Finally in section $5$ we present a discussion of the results
and present our conclusions.

In what follows we will use natural units ($\hbar=c=k_{B}=8\pi G=1$) and the signature $(+,-,-,-)$. 

\section{Basic Equations} \label{Equazioni e convenzioni}
The general action for a fourth order theory of gravity is:
\begin{eqnarray}
\mathcal{A}=\int d^{4}x\sqrt{-g}[f(R)+L_{M}],
\end{eqnarray}
where $f(R)$ is a function of Ricci scalar $R$ and $L_{M}$ is the
standard matter  Lagrangian density. Varying this action with respect
to the metric gives
\begin{eqnarray}\nonumber
\fl G_{\mu\nu}&=&T^{ToT}_{\mu\nu}=T^{M}_{\mu\nu}+T^{R}_{\mu\nu}
=\frac{1}{f'(R)}\tilde{T}^{M}_{\mu\nu}\,\\
\fl &+&\frac{1}{f^{\prime} (R)}\Big(  \frac{1}{2}g_{\mu\nu}\left[f(R)-R f^{\prime}(R)\right]
+f^{\prime}(R)^{;\alpha\beta}\left(g_{\alpha\mu}g_{\beta\nu}-g_{\alpha\beta} g_{\mu\nu}\right) \Big)\;,
\end{eqnarray}
where the prime denotes the derivative with respect to R,
$T^{M}_{\mu\nu}$ is the  effective stress-energy tensor for standard
matter, which is assumed to be a perfect fluid and $T^{R}_{\mu\nu}$
is the stress-energy tensor of the curvature {\em fluid}  which
represents an additional source term of purely geometrical origin
\cite{8}. By assuming $f(R)=exp\,(-{R}/{\Lambda})$ we obtain the
field equations:
\begin{eqnarray}
\fl G_{\mu\nu}&=&-\Lambda e^{R/\Lambda}
\tilde{T}^{M}_{\mu\nu}\,\\ \fl &-&e^{-R/\Lambda} \left[
\frac{1}{2}g_{\mu\nu}(\Lambda+R)+\frac{1}{\Lambda^{2}}(R^{;\alpha}R^{;\beta}+\Lambda
R^{;\alpha\beta}\small)
\left(g_{\alpha\mu}g_{\beta\nu}-g_{\alpha\beta}
g_{\mu\nu}\right)\right]\,.
\end{eqnarray}
In the case of the Friedmann-Lemaitre-Robertson-Walker (FLRW) metric, the above
equations reduce to:
\begin{equation}
\eqalign{\qquad\qquad\qquad H^{2}+\frac{k}{a^{2}}-\frac{H}{\Lambda}\dot{R}+\frac{R}{6}+
\left(\frac{\Lambda}{6}+\frac{\Lambda \rho}{3e^{-R/\Lambda}}\right)&=0,\label{CE1}\\
2\frac{\ddot{a}}{a}+\frac{R}{3}-\frac{H}{\Lambda}\dot{R}+\frac{1}{\Lambda^{2}}\dot{R}^{2}
-\frac{1}{\Lambda}\ddot{R}-\frac{\Lambda\rho}{3e^{-R/\Lambda}}(1+3w)+\frac{\Lambda}{3}&=0,}
\nonumber
\end{equation}
with,
\begin{eqnarray}
R=-6\left(\frac{\ddot{a}}{a}+H^{2}+\frac{k}{a^{2}}\right)\;,
\label{CE2}
\end{eqnarray}
where $H=\frac{\dot{a}}{a}$ is the usual Hubble parameter and $k$ is
the spatial curvature.  The Bianchi identities applied to the total
stress-energy tensor $T_{\mu\nu}^{TOT}$ lead to the energy
conservation equation for standard matter \cite{a}:
\begin{eqnarray}\label{cons matter}
\dot{\rho}+3H\rho(1+w)=0\;.
\end{eqnarray}

\section{\textbf{The vacuum case}}

In the vacuum case $(\rho=0)$ equation \rf{CE1} and \rf{CE2} can be
written as a closed system  of first order differential equations
using the dimensionless variables,
\begin{eqnarray}\label{variables}
x=\frac{\dot{R}}{\Lambda H},\;\;\;\;y=\frac{R}{6H^{2}},\;\;\;\;z=
\frac{\Lambda}{6H^{2}},\;\;\;\;K =\frac{k}{a^{2}H^{2}}\;.
\end{eqnarray}
It is clear that the variables $y$ and $z$ are a measure of the
expansion normalized Ricci curvature and the cosmological constant
respectively, $K$ is the spatial curvature parameter of the Friedmann
model, while $x$ is a measure of the time rate of change of Ricci
curvature. The evolution equations for the variables \rf{variables}
are given by
\begin{eqnarray} \label{gensys}\nonumber
x'&=&2z+2K-2+x(1+x+y+K)\;,\\  \nonumber
y' &=&x z+2y(2+y+K)\;,\\
z'&=&2z(2+y+K)\;,\\
K'&=&2K(y+1+K)\;, \nonumber
\end{eqnarray}
where the prime represents the derivatives with respect to the time
variable $\mathcal{N}=\ln\,a$. This system is completed with the Friedmann constraint,
\begin{eqnarray}\label{Gauss vacuum}
1+K+y+z-x=0\;,
\end{eqnarray}
which defines a hyperplane in the total
phase space of the system. Consequently, all solutions of the
dynamical system will be located in a non-compact submanifold of the
phase space associated with \rf{gensys}. The time derivative of
\rf{Gauss vacuum} is nothing other than the Raychaudhuri equation.

\subsection{Finite analysis}

The dimensionality of the state space of the system \rf{gensys} can
be reduced by eliminating any one of the four variables using the
constraint equation \rf{Gauss vacuum}. If we choose to eliminate the $x$ the
dynamical equations become
\begin{eqnarray}\label{dynsys red vacuum}\nonumber
y'&=&y(4+2K+2y+z)+z(1+K+z)\;,\\
z'&=&2z(2+K+y),\\ \nonumber
K'&=&2K(1+K+y)\;,
\end{eqnarray}
and the number of invariant submanifolds is maximized, making the analysis much easier. 
It is clear from the above equations that $z=0$ (corresponding to the $(K-y)$ plane) is an invariant
submanifold.
 Specifically, If we choose $z\neq 0$ as an initial condition for
our cosmological evolution, any general orbit can only approach
$z=0$ asymptotically. This implies that no orbit crosses the $(K-y)$
plane and consequently no global attractor can exist.

Setting $K^{\prime}=0$, $y^{\prime}=0$, $z^{\prime}=0$, we obtain four fixed points.
We can obtain exact cosmological solutions at these points 
using the equation,
\begin{eqnarray}\label{Rayfxpt}
\dot{H}=-(y+K+2)H^{2}\;.
\end{eqnarray}
In fact, at any fixed point, equation \rf{Rayfxpt} reduces to
\begin{eqnarray}\label{Rayfxptsimpl}
\dot{H}=-\frac{1}{\alpha} H^{2}\;,
\end{eqnarray}
where
\begin{equation}
\alpha=(y_{*}+K_{*}+2)^{-1}\;,~~\alpha\neq0
\end{equation}
and the quantities $X_*$ are meant to be calculated at the fixed
point. Equation \rf{Rayfxpt} applies to both the matter and vacuum
cases and describes a general power law evolution of the scale
factor. In addition integrating with respect to time we obtain
\begin{eqnarray}
a=a_{o}(t-t_{o})^{\alpha}\;.
\end{eqnarray}
This means that by finding the value
of $\alpha$ at a given fixed point, we can obtain the solutions
associated with it using equation \rf{Rayfxpt}.

In this way, points $\mathcal{A}_{v}$ and $\mathcal{B}_{v}$ (see Table \ref{fiefixpoint}) are found to represent Milne and power-law evolutions respectively. However, by  direct substitution
into the cosmological equations it can be shown that these fixed
points cannot be considered as physical points, which means that
although we can choose initial conditions as close as we want to
these points, the cosmology will never evolve according to these
solutions.

For the points $\mathcal{C}_{v}$ and $\mathcal{D}_{v}$ we have $y_{*}+K_{*}+2= 0$. In this case
\begin{eqnarray}
\dot H =0,
\end{eqnarray}
which implies
\begin{eqnarray}
a=a_{o}e^{\gamma(t-t_{o})}.
\end{eqnarray}
The value of the constant $\gamma$ can be obtained by direct
substitution  into equations \rf{CE1} and \rf{CE2}. For both $\mathcal{C}_{v}$ and
$\mathcal{D}_{v}$ we obtain
\begin{eqnarray}
\gamma=\pm \sqrt{\frac{\Lambda}{6}}\;,
\end{eqnarray}
so they represent an exponential evolution. The contracting or
expanding nature of this solution depends on the direction of
approach of the orbits with respect to the hypersurface $y+K+2=0$. This hypersurface
divides the phase space in two hypervolumes characterized by a
contracting or expanding evolution. In particular, for $y<-K-2$ the
orbits describe a contracting universe, while for  $y>-K-2$
they represent an expanding one.

The stability of the hyperbolic fixed points $\mathcal{A}_{v}$, $\mathcal{B}_{v}$ and $\mathcal{D}_{v}$ is
obtained by using Hartman-Grobman theorem, while for the point $\mathcal{C}_{v}$,
which is non-hyperbolic,  we use the local center manifold theorem
to find it's stability. A brief review of this approach in presented
in the Appendix.

In our case, using the transformation
\begin{eqnarray}\nonumber \label{tr}
y=u-2v-m\;,\\
z=4m\;,\\ \nonumber
k=v\ \nonumber
\end{eqnarray}
the system \rf{dynsys red vacuum} can be written in the
diagonal form
\begin{eqnarray}
\dot{u}&=&-4u+12m^2+2mu+2u^2-4mv-2uv\;,\\
\dot{v}&=&-2v-2mv+2uv-2v^2\;,\\
\dot{m}&=&-2m^2+2mu-2mv
\end{eqnarray}
near the non-hyperbolic fixed point $\mathcal{C}_{v}$.
By substituting the expansions
\begin{eqnarray}
h1(m)&=am^2+bm^3 +O(m^4)\;, \\
h2(m)&=cm^2+dm^3 +O(m^4)
\end{eqnarray}
into equations \rf{EqW0} and \rf{EqW0appr} and then solving for
$a$, $b$, $c$ and $d$, we obtain
\begin{eqnarray}
h1(m)=3m^2+\frac{9}{2}m^3+O(m^4),\qquad h2(m)=O(m^4)\;.
\end{eqnarray}
Substituting this result into equation \rf{EqW0} then yields
\begin{eqnarray}
\dot{m}=-2m^2+O(m^3)
\end{eqnarray}
on the center manifold $W^{c}(\textbf{0})$, near the point $\mathcal{C}_{v}$. This
implies that the point $\mathcal{C}_{v}$ is a saddle-node i.e. it behaves like a saddle or an attractor depending on the direction
from which the orbit approaches. The local phase portrait
in the neighborhood of $\mathcal{C}_{v}$ is shown in Figure 1 .

If one considers now the transformation (\ref{tr}) one realizes
that $m\propto z$, so that $\mathcal{C}_{v}$ is an attractor for $z>0$ and a
saddle for $z<0$. This also clear from Figure \ref{z-y} in which the
invariant submanifold $z-y$ is depicted.

Finally, it is useful to derive an expression for the deceleration parameter $q$ in terms of
the dynamical variables:
\begin{eqnarray} \label{q vacuum}
q=-\frac{\dot H}{H^{2}}-1 = -(y+K+1)\;.
\end{eqnarray}
It follows that  $q > 0 \Rightarrow (y+K+1)< 0$. This condition is
satisfied  only for the point $\mathcal{C}_{v}$ as expected by looking at the
solution associated with this fixed point. In Figure \ref{K-y} we
give the location of the $q=0$ plane relative to the fixed points
$\mathcal{A}_{v}$, $\mathcal{C}_{v}$ and $\mathcal{B}_{v}$.
\begin{table}
\caption{ Coordinates of the fixed points, eigenvalues, stability
and solutions for $exp\,(-\frac{R}{\Lambda})$-gravity in vacuum.}\label{fiefixpoint}
\begin{center}
\begin{tabular}{lllll}
\br
{\small Point}&{\small Coordinates(y,z,K)}&{\small Eigenvalues}&{\small Stability}&{\small Solution}\\
\mr
$\mathcal{A}_{v}$ & $[0,0,0]$    & $[2,4,4]$          &{\small repeller} &$ a=a_{o}(t-t_{o})$  \\
$\mathcal{B}_{v}$ & $[0,0,-1]$  &$ [-2,2,2]$      &{\small Saddle} & $a=a_{o}(t-t_{o})^{\frac{1}{2}}$  \\
$\mathcal{C}_{v}$ & $[-2,0,0] $ &$[-4,-2,0]$       &{\small Saddle-node} &$ a=a_{o}e^{\gamma(t-t_{o})}$ \\
$\mathcal{D}_{v}$ &$[-2,1,0]$ &$[-\frac{(3+\sqrt{17})}{2},-2,\frac{(3+\sqrt{17})}{2}]$ & {\small Saddle} & $a=a_{o}e^{\gamma(t-t_{o})}$ \\
\br
\end{tabular}
\end{center}
\end{table}


\begin{figure}
\begin{center}
 \includegraphics[width=9cm,height=9cm]{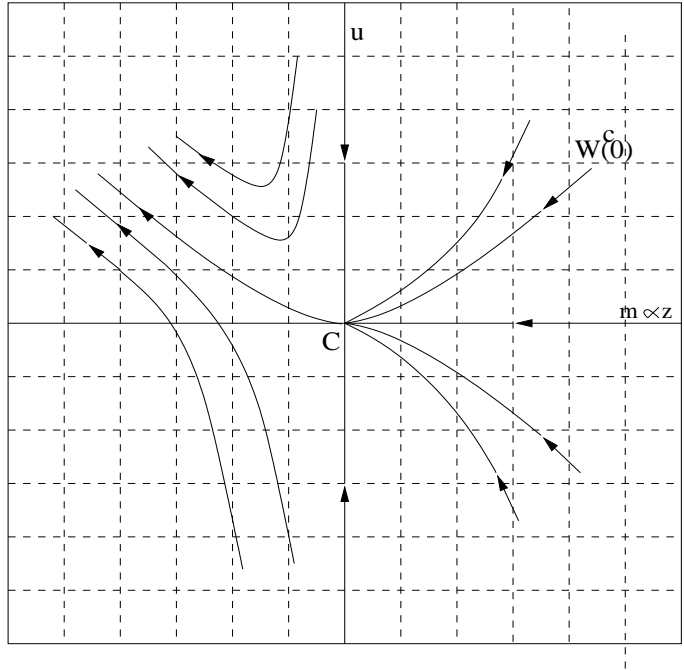}
\caption{The phase portrait for the system $(12)$ in the neighbourhood of the fixed point $\mathcal{C}$ for $exp\,(-R/\Lambda)$-gravity in vacuum.}
\label{pointC}
\end{center}
\end{figure}

\begin{figure}
\begin{center}
 \includegraphics[width=9cm,height=9cm]{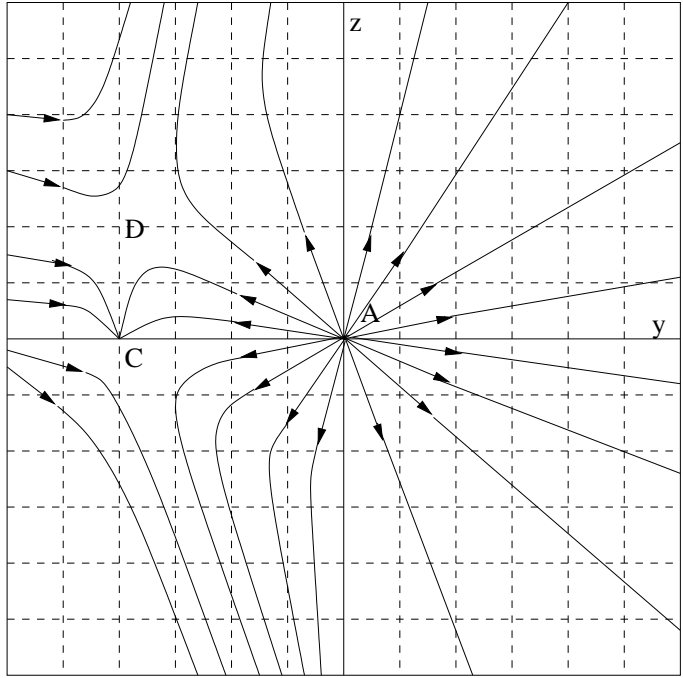}
\caption{The phase space of the invariant submanifold $z-y$ for $exp\,(-R/{\Lambda})$-gravity in vacuum.}
\label{z-y}
\end{center}
\end{figure}

\begin{figure}
\begin{center}
 \includegraphics[width=9cm,height=9cm]{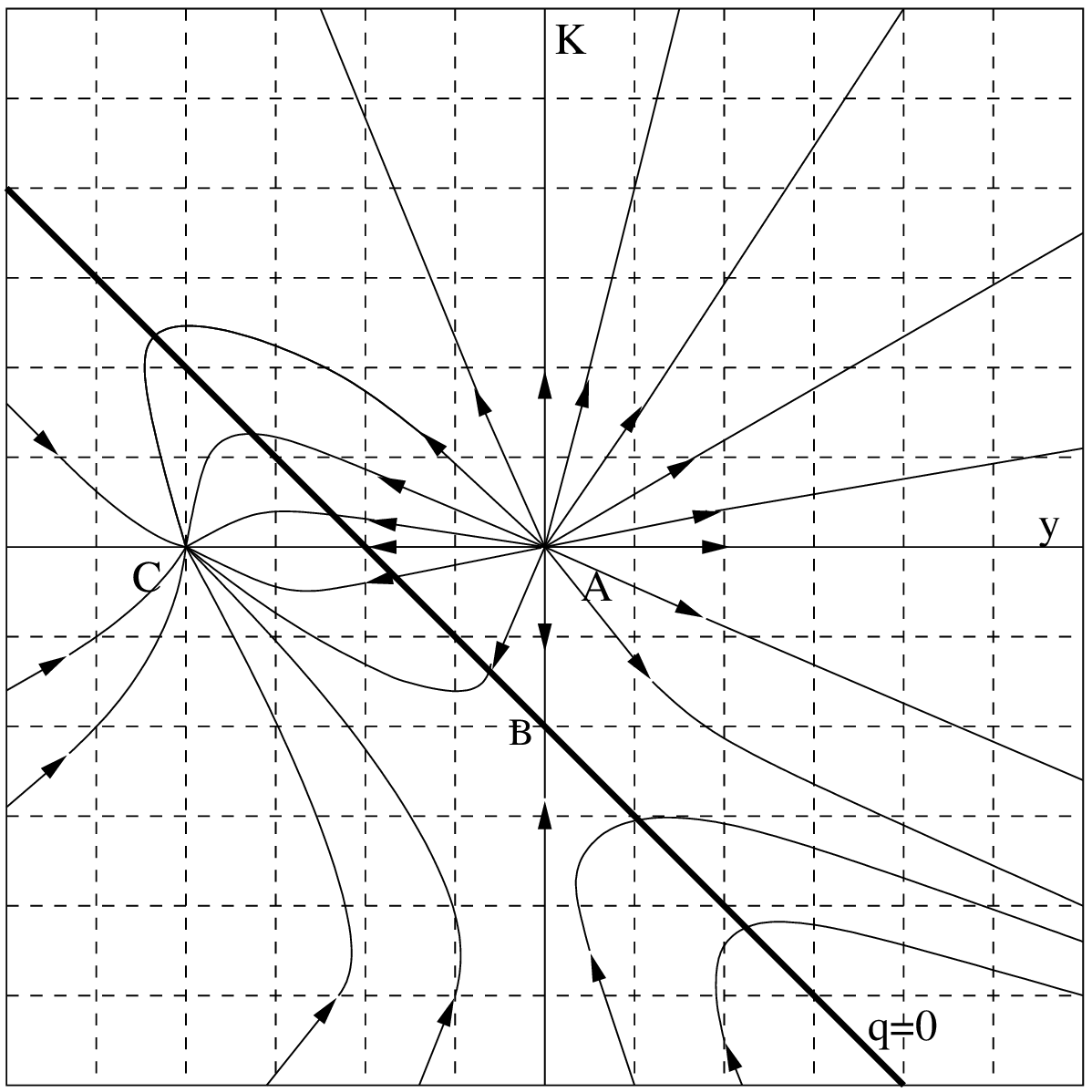}
\caption{The location of the $q=0$ plane relative to the fixed points $\mathcal{A}_{v}$, $\mathcal{C}_{v}$ and $\mathcal{B}_{v}$ for $exp\,(-R/{\Lambda})$-gravity in vacuum.}
\label{K-y}
\end{center}
\end{figure}

\subsection {Asymptotic analysis }
In this section we will determine the  fixed points at infinity and
study their stability. To simplify the asymptotic analysis we need
to compactify the phase space using the Poincar\'{e} method. 
Transforming to polar coordinates $(r( N),\theta( N),\phi N))$:
\begin{eqnarray*}
z\rightarrow r\, \,\cos\theta , \;\;  K\rightarrow r\,
\,\sin\theta \;  \cos \phi, \;\; y\rightarrow r \,
\,\sin\theta\; \sin\phi
\end{eqnarray*}
and substituting $r\rightarrow \frac{R}{1-R}$, the regime
$r\rightarrow \infty$ corresponds to $R\rightarrow 1$. Using
this coordinate transformation and taking the limit
$R\rightarrow \infty$, the system $(13)$ can be written as
\begin{eqnarray}\nonumber
R'\rightarrow
\frac{1}{4}&\Big(& 8\cos(\phi)\sin(\theta)^3-\sin(\phi)\left(-7\sin(\theta)+\sin(3\theta)\right)
\\&+&8\cos(\theta)^2
sin(\theta)
(\chi)+4\cos(\theta)\sin(\theta)^2\sin(\phi)
\left(\chi\right)\Big)\;,
\end{eqnarray}
\begin{eqnarray}
R \theta'\rightarrow
\frac{-\cos(\theta)^2\sin(\phi)\left(\cos(\theta)+\sin(\theta)\left(\chi\right)\right)}{R-1}\;,
\end{eqnarray}
\begin{eqnarray}
R\phi' \rightarrow
\frac{\cos(\phi)\cot(\theta)\left(\chi+\cot(\theta)\right)}{R-1}\;,
\end{eqnarray}
where $\chi=\cos(\phi)+\sin(\phi)$ . Since equation $(30)$ does not
depend on the coordinate $R$, we can find the fixed points of the
above system using equations $(31)$ and $(32)$ only. The results are
shown in Table \ref{asyfixpoint}.

Let us now  derive the solution for the first fixed point,  which 
corresponds to $K\rightarrow-\infty, z\rightarrow\infty$. In
these limits the first equation of the system $(13)$ reduces to
\begin{eqnarray}
\dot{K}=2K
\end{eqnarray}
and the equation for $\dot{H}$ becomes
\begin{eqnarray}
\dot{H}=KH^2\;.
\end{eqnarray}
Integrating equation $(33)$ we obtain
\begin{eqnarray}
K=-\frac{1}{2(N-N_{\infty})}\;.
\end{eqnarray}
Substituting $K$ back into equation $(34)$ and solving for $(N-N_{\infty})$, we obtain
\begin{eqnarray}
(N-N_{\infty})=[c_{1}\pm\frac{3}{2}c_{o}(t-t_{o})]^{\frac{2}{3}}\;.
\end{eqnarray}
The same procedure can be used to obtain solutions for the other
asymptotic points.  On the fixed circle $\mathcal{I}_{\infty}$ we have three possible
solutions corresponding to whether the sign of $K$ and $y$ are the
same or different.

The stability of the fixed points in Table \ref{asyfixpoint} are found by expanding
equations  $(31)$ and $(32)$ up to second order and then
applying the center manifold theorem to the resulting system (see
Table \ref{Stab}).

For the fixed circle $\mathcal{I}_{\infty}$, the stability depends on the value of the
angle $\phi$:
\begin{center}
\begin{tabular} {lll}
$-\pi/4< \phi <\pi/2$   $\quad$  & {\small Stable},\\
$\pi/2< \phi <3\pi/4$    $\quad$  & {\small Unstable},\\
$3\pi/4< \phi <3\pi/2$   $\quad$  & {\small Stable},\\
$3\pi/2< \phi <7\pi/4$    $\quad$ & {\small Unstable}.\\
\end{tabular}
\end{center}
In the next section we will see how the introduction of matter
modifies the picture we obtained in the vacuum case.

\begin{table}\caption{  Coordinates, eigenvalues and the Stability of the fixed points
in the asymptotic regime for $exp\,(-\frac{R}{\Lambda})$ gravity in
vacuum.}\label{asyfixpoint}
\begin{center}
\begin{tabular}{lllll}
\br
{\small Point} & $(\theta,\phi)$ & {\small Eigenvalues} & {\small Coordinates}& {\small Solution}  \\
\mr
$\mathcal{A}_v\infty$& $[\frac{-\pi}{4},0]$ & $[2,0] $&$|K|,|z|\rightarrow\infty$ $y \rightarrow$ 0 & $(N-N_{\infty})=[c_{1}\pm\frac{3c_{o}}{2}(t-t_{o})]^{\frac{2}{3}}$\\
$\mathcal{B}_v\infty$ & $[\frac{\pi}{4},-\pi]$ & $[2,0] $&$|K|,|z|\rightarrow\infty$ $y \rightarrow$ 0 &$(N-N_{\infty})=[c_{1}\pm\frac{3c_{o}}{2}(t-t_{o})]^{\frac{2}{3}}$ \\
$\mathcal{E}_v\infty$& $[\frac{-3\pi}{4},\frac{-\pi}{2}]$ &$[\frac{-1}{2\sqrt{2}},0]$&  $|y|,|z|\rightarrow\infty$ $K\rightarrow 0$ &$(N-N_{\infty})=[c_{1}\pm\frac{c_{o}}{2}(t-t_{o})]^{2}$ \\
$\mathcal{F}_v\infty$ & $[\frac{\pi}{4},\frac{-\pi}{2}]$ & $[\frac{1}{2\sqrt{2}},0]$ &$|y|,|z|\rightarrow\infty$ $K\rightarrow 0$&$(N-N_{\infty})=[c_{1}\pm\frac{3c_{o}}{2}(t-t_{o})]^{\frac{2}{3}}$ \\
$\mathcal{G}_v\infty$ & $[\frac{-\pi}{4},\frac{\pi}{2}]$ & $[\frac{-1}{2\sqrt{2}},0]$&$|y|,|z|\rightarrow\infty$ $K\rightarrow 0$&$(N-N_{\infty})=[c_{1}\pm\frac{3c_{o}}{2}(t-t_{o})]^{\frac{2}{3}}$ \\
$\mathcal{H}_v\infty$ & $[\frac{3\pi}{4},\frac{\pi}{2}]$ & $[\frac{1}{2\sqrt{2}},0]$ & $|y|,|z|\rightarrow\infty$ $z\rightarrow$ 0& $(N-N_{\infty})=[c_{1}\pm\frac{c_{o}}{2}(t-t_{o})]^{2}$\\
$\mathcal{I}_v\infty$ & $[\frac{\pi}{2},\phi]$ & $[0,-\chi\cos(\phi)]$ &$y\rightarrow
\pm\infty$
$K\rightarrow \mp\infty$& $a=a_{0}{\small exp}[\gamma(t-t_{0})]$\\
 &  &
&$y\rightarrow +\infty$
$K\rightarrow +\infty$& $(N-N_{\infty})=[c_{1}\pm\frac{c_{o}}{2}(t-t_{o})]^{2}$\\
 &  &
 &$y\rightarrow -\infty$
$K\rightarrow -\infty$& $(N-N_{\infty})=[c_{1}\pm\frac{3c_{o}}{2}(t-t_{o})]^{\frac{2}{3}}$\\
\br
\end{tabular}
\end{center}
\end{table}

\begin{table}\caption{Coordinates of the asymptotic  fixed points of
 $exp\,(-R/{\Lambda})$-gravity in vacuum and their stability.}\label{Stab}
\begin{center}
\begin{tabular} {lll}
\br
{\small Point} & {\small Coordinates(y,z,K)}   & {\small Stability} \\
\mr
$\mathcal{A}_\infty$ &$[-\pi/4,0] $   \hspace{.8in}       & {\small Unstable}\\
$\mathcal{B}_\infty$ &$[\pi/4,-\pi ]$   \hspace{.8in}     & {\small Stable}\\
$\mathcal{E}_\infty$ &$[-3\pi/4,-\pi/2] $ \hspace{.8in}   & {\small Stable}\\
$\mathcal{F}_\infty$ &$[\pi/4,-\pi/2]$  \hspace{.8in}     & {\small Stable}\\
$\mathcal{G}_\infty$ &$[-\pi/4,\pi/2]$  \hspace{.8in}     & {\small Stable}\\
$\mathcal{H}_\infty$ &$[3\pi/4,\pi/2] $  \hspace{.8in}    & {\small Stable}\\
\br
\end{tabular}
\end{center}
\end{table}


\section{\textbf{The matter case}}
In this case we can use the same dynamical variables we used for
the vacuum case together with one additional variable $D$, that is
related to the matter energy density:
\begin{eqnarray}
x=\frac{\dot{R}}{\Lambda
H},\,\;\;y=\frac{R}{6H^{2}},\;\;\,z=\frac{\Lambda}{6H^{2}},\,\;
\;K=\frac{k}{a^{2}H^{2}},\,\;\;D=\frac{\Lambda\rho}{3H^{2}e^{R/\Lambda}}\;.\end{eqnarray}
The definition of the variables reveals that not all of the phase
space corresponds to physical situations. This becomes  clear if we
divide $D$ by $z$. We obtain
\begin{eqnarray}
\frac{D}{z}=2\,\rho\,\,exp\left(-\frac{R}{\Lambda}\right)\;,
\end{eqnarray}
which has the same sign of $\rho$. This means that the sectors in
the phase space for which the sign of $D$ is different from the sign
of $z$ contain orbits in which standard matter violates the weak
energy condition $\rho>0$, and have to be discarded as not  
physical. As we will see this affect the sets of possible orbits for
this model.

Following the same procedure we used in the vacuum case, we
obtain an autonomous system equivalent to the cosmological
equations with non-zero matter density:
\begin{eqnarray}\label{dynsys red matter}
\nonumber x'&=&2+2z+2K+x(1-x+y+K)-D(1+3w)\;,\\ \nonumber
y'&=&x z+2y(2+y+K)\;,\\
z'&=&2z(2+y+K)\;,\\ \nonumber
K'&=&2K(y+1+K)\;,\\ \nonumber
D'&=&D(1-3w+2y+2K-x)\;,\nonumber
\end{eqnarray}
together with the constraint equation
\begin{eqnarray}\label{Gauss matter}
1+K+x+y-z-D=0\;,
\end{eqnarray}
where the prime again denotes the derivative with respect to the logarithmic time 
variable $\mathcal{N}$.
\subsection{Finite analysis}
The system \rf{dynsys red matter} can be further simplified, by
eliminating $x$ using the constraint equation \rf{Gauss matter}:
\begin{eqnarray}\nonumber\label{mattereq}
y'&=&y(4+2K+2y+z)+z(1+K+D++z)\;,\\
K'&=&2K(1+K+y)\;,\\\nonumber
z'&=&2z(2+y+K)\;,\\
D'&=&D(2-3w+3K+D+3y+z)\;.\nonumber
\end{eqnarray}
We have three invariant submanifolds: $K=0$, $z=0$
and $D=0$, so in this case also no global attractor can exist.
Setting $K^{\prime}=0$, $y^{\prime}=0$, $z^{\prime}=0$ and $D^{\prime}=0$
we obtain seven fixed points.

As in the vacuum case, we can use  the coordinates of these fixed
points and equation \rf{Rayfxpt} to find the behaviour of the scale
factor at these points. In addition, the behaviour of 
the energy density $\rho$ can be obtained from equation \rf{cons matter}, 
which at a fixed point reads
\begin{equation}\label{cons matter2}
\frac{\dot{\rho}}{\rho}=-3(1+w)\frac{\alpha}{t}\;,
\end{equation}
where $\alpha$ is defined by \rf{Rayfxptsimpl}. However, direct
substitution in the cosmological equations reveals that all the
fixed points correspond to vacuum states.

Points $\mathcal{A}_{m}$ and $\mathcal{B}_{m}$ are found to represent Milne solutions while $\mathcal{C}_{m}$
and $\mathcal{D}_{m}$ represent a power law evolution. For points $\mathcal{E}_{m}$, $\mathcal{F}_{m}$ and $\mathcal{G}_{m}$ we find that $\dot{H}=0$, which means that these points
represent Einstein-de Sitter solutions. The exact solutions at these fixed points
are summarized in Table \ref{fix point matt}.

As in the vacuum case, we use the Hartman-Grobman theorem 
together with the center manifold theorem to analyze the stability
of all the fixed points. The results are shown in Table \ref{stabfix
point matt}.

Equation \rf{q vacuum}, which relates the deceleration parameter to
the dynamical variables generalizes in the matter case to the hyperplane,
\begin{eqnarray}
q=-\frac{\dot H}{H^{2}}-1 = -(y+K+1).
\end{eqnarray}

\begin{table}\caption{Coordinates of the fixed points, the eigenvalues, and solutions
for $exp\,(-R/{\Lambda})$-gravity in the matter case.} \label{fix
point matt}
\begin{center}
\begin{tabular}{lllll}
\br
{\small Point} & {\small Coordinates(y,z,K,D) } & {\small Eigenvalues} & {\small Solution } \\
\mr
$\mathcal{A}_{m}$ & $[0,0,0,0]$     & $[2-3 w, 2, 4,4]$ &$ a=a_{o}(t-t_{o})^{\frac{1}{2}}$ \\
$\mathcal{B}_{m}$  &$[ 0, 0,  -1,0]$ &$[2,2,-2,-(1+3w)]$&$ a=a_{o}(t-t_{o})$ \\
$\mathcal{C}_{m}$  &$[ 0, 0, 0,3w-2] $&$[3 w - 2, 2, 4,4]$&$ a=a_{o}(t-t_{o})^{\frac{1}{2}}$  \\
$\mathcal{D}_{m}$  &$[0, 0, -1, 3 w +1]$&$[2,2,-2,(1+3w)]$&$ a=a_{o}(t-t_{o})$ \\
$\mathcal{E}_{m}$  &$[-2,1, 0 , 0] $
&$[-\frac{\sqrt{17}+3}{2},\frac{\sqrt{17}-3}{2},-2,-3-3w]$ &
$a=a_{o}e^{\gamma(t-t_{o})}$\\
$\mathcal{F}_{m}$  &$[ -2,  0, 0,  0] $&$[- 2, - 4, - 3 w - 4, 0]$ &$a=a_{o}e^{\gamma(t-t_{o})}$\\
$\mathcal{G}_{m}$  &$[-2,  0,  0,  3w+4] $&$[3 w + 4, - 2, - 4, 0]$ &$a=a_{o}e^{\gamma(t-t_{o})}$\\
\br
\end{tabular}
\end{center}
\end{table}

\begin{table}\caption{Stability of the fixed points for $exp\,(-R{\Lambda})$-gravity
in the matter case.}\label{stabfix point matt}
\begin{center}
\begin{tabular}{lllll}
\br
{\small Point} & $w=0$ & $0<w<\frac{1}{3}$ & $w=\frac{1}{3}$  \\
\mr
$\mathcal{A}_{m}$ & {\small Repeller}    & {\small Repeller}    & {\small Repeller} \\
$\mathcal{B}_{m}$ & {\small Saddle}      &   {\small Saddle}    &{\small Saddle}\\
$\mathcal{C}_{m}$ & {\small Saddle}      & {\small Saddle}      & {\small Saddle} \\
$\mathcal{D}_{m}$ & {\small Saddle}      & {\small Saddle}      & {\small Saddle} \\
$\mathcal{E}_{m}$ & {\small Saddle}      & {\small Saddle}      & {\small Saddle} \\
$\mathcal{F}_{m}$ & {\small Saddle-node} & {\small Saddle-node} & {\small Saddle-node} \\
$\mathcal{G}_{m}$ & {\small Saddle-node} & {\small Saddle-node} & {\small Saddle-node} \\
\br
{\small Point} & $\frac{1}{3}<w<\frac{2}{3}$ & $w=\frac{2}{3}$ & $\frac{2}{3}<w<1$ \\
\mr
$\mathcal{A}_{m}$ &  {\small Repeller}   & {\small Saddle-node} &{\small Saddle} \\
$\mathcal{B}_{m}$ &  {\small Saddlee}    & {\small Saddle}      & {\small Saddle} \\
$\mathcal{C}_{m}$ &  {\small Saddle}     & {\small Saddle-node} & {\small Repeller}\\
$\mathcal{D}_{m}$ &  {\small Saddle}     & {\small Saddle}      & {\small Saddle}\\
$\mathcal{E}_{m}$ &  {\small Saddle}     & {\small Saddle}      & {\small Saddle} \\
$\mathcal{F}_{m}$ & {\small Saddle-node} & {\small Saddle-node} & {\small Saddle-node} \\
$\mathcal{G}_{m}$ & {\small Saddle-node} & {\small Saddle-node} & {\small Saddle-node} \\
\br
\end{tabular}
\end{center}
\end{table}

\subsection { Asymptotic analysis }

We complete the analysis for the matter case by investigating the
asymptotic behavior of the system \rf{mattereq}. In order to achieve this we
compactify  the phase space by transforming to 4-D  polar coordinates. The
transformation equations are
\begin{eqnarray*}
D\rightarrow r\,\, \cos\delta,  \;\; z\rightarrow r \,\,\sin\delta\,\,\, \cos\theta,\;\;
 x\rightarrow r\,\, \sin\delta\,\,\, \sin\theta\,\,\, \cos\phi, \;\;\\
y\rightarrow r\,\, sin\theta \,\,\,\sin\delta\,\,\, \sin\phi\;,
\end{eqnarray*}
where $r\in[0,\infty[,\,\delta\in[0,\pi],\,\theta\in[0,\pi],\, $ and $\phi\in[0,2\pi]$.
We then transform the radial coordinate $r\rightarrow\frac{R}{1-R}$ and in the limit $R\rightarrow 1$,
the system $(40)$ reduces to
\begin{eqnarray}\nonumber
\fl R' \rightarrow
\cos(\delta)^3&+&\cos(\delta)\cos(\theta)\sin(\delta)^{2}\sin(\theta)\sin(\phi)
+\cos(\delta)^{2}\sin(\delta)(\cos(\theta)\\ \fl &+&3\sin(\theta)\chi)
+\sin(\delta)^{3}\sin(\theta)\Big(\cos(\phi)(2+\varphi)\\\nonumber\fl &
+&\sin(\phi)(3\cos(\theta)^{2}+2\sin(\theta)^{2}+\varphi)\Big)\;,
\end{eqnarray}
\begin{eqnarray}\nonumber
\fl R\delta'\rightarrow
\frac{\sin(\delta)\cos(\delta)}{8(R-1)}&\Big[&8\cos(\delta)(\varphi-1)
-\sin(\delta)\Big( \cos(3\theta)+8\cos(\phi)\sin(\theta)\\\nonumber
\fl &+&8\sin(\theta)^{3}\sin(\phi)+\cos(\theta)(7+4\sin(\theta)^{2}(\cos(2\phi)\\&-&\sin(2\phi))) \Big) \Big]\;,
\end{eqnarray}
\begin{eqnarray}\nonumber
\fl R\theta'\rightarrow \frac{\cos(\theta)^{2}}{2(R-1)}
&\Big[&2\cos(\delta)\sin(\phi)+\sin(\delta)(2\cos(\theta)\sin(\phi)
\\ \fl &+&\sin(\theta)(1-\cos(2\phi)+\sin(2\phi))) \Big]\;,
\end{eqnarray}
\begin{eqnarray}
\fl R\phi'\rightarrow \frac{\cos(\phi)(\cos(\delta)\cot(\theta)
+\cos(\theta)\sin(\delta)(\sin(\phi)+\cos(\phi)+\cot(\theta)))}{R-1}\;,
\end{eqnarray}
where $\varphi=\cos(\theta)\sin(\theta)\sin(\phi)$. Notice that the
first equation of the previous system does not depend on $R$, which
means that the fixed points of this system can be determined by
the angular equations alone. The solutions at the fixed points 
can then be obtained by following the same procedure we
used in the vacuum case. All the results are summarized in Table
\ref{asy fix point matt}. The stability of the first six fixed points
are shown in Table \ref{tab asy matter}.

As before, we use the center manifold theorem to determine the
stability of these fixed points. In this case, because the
eigenvalues have a double zero, the coordinate which correspond to
the non-zero eigenvalue are approximated by the function
\begin{eqnarray}
\textbf{h}(\textbf{x}_1,\textbf{x}_2)=a\,\textbf{x}_1^2+b\,\textbf{x}_1^2 \textbf{x}_2^2+c\,\textbf{x}_2^2+0(|\textbf{x}|^3),
\end{eqnarray}
where $\textbf{x}_1$ and $\textbf{x}_2$ are the coordinates that correspond to the zero
eigenvalues (see the Appendix for details). The stability of the point $\mathcal{M}_m\infty$ depends on the
angle $\phi$: for $-\frac{\pi}{4}>\phi>\frac{3\pi}{4}$, it
is unstable, otherwise it is stable. Finally the point $\mathcal{O}_m\infty$ is unstable for all values of
$\phi$.

\begin{table}\caption{ Coordinates, eigenvalues, and the solutions for fixed
points in the asymptotic regime for the $exp\,(-R/{\Lambda})$
gravity in matter case.}\label{asy fix point matt}
\begin{center}
\begin{tabular}{llllll}
\br
{\small Point}  &$(\delta,\theta,\phi)$ & {\small Eigenvalues} &{\small Solution}\\
\mr
$\mathcal{A}_m\infty$ & $[-\frac{\pi}{2},-\frac{3\pi}{2},-\frac{\pi}{2}]$ &$[1,0,0]$&$(N-N_{\infty})=[c_{1}\pm\frac{3c_{o}}{2}(t-t_{o})]^\frac{2}{3}$\\
$\mathcal{B}_m\infty$ & $[\frac{\pi}{2},-\frac{3\pi}{2},-\frac{\pi}{2}]$ & $[-1,0,0]$&$(N-N_{\infty})=[c_{1}\pm\frac{c_{o}}{2}(t-t_{o})]^{2}$ \\
$\mathcal{C}_m\infty$ & $[-\frac{\pi}{2},\frac{\pi}{4},-\frac{\pi}{2}]$ & $[-\frac{1}{\sqrt{2}},0,0]$&$(N-N_{\infty})=[c_{1}\pm\frac{c_{o}}{2}(t-t_{o})]^{2}$ \\
$\mathcal{D}_m\infty$ & $[-\frac{3\pi}{4},\frac{\pi}{2},-\frac{\pi}{2}]$ & $[-\frac{1}{\sqrt{2}},0,0]$ & $(N-N_{\infty})=[c_{1}\pm\frac{c_{o}}{2}(t-t_{o})]^{2}$ \\
$\mathcal{E}_m\infty$ & $[\frac{\pi}{2},\frac{\pi}{4},-\frac{\pi}{2}]$
&$[\frac{1}{\sqrt{2}},0,0]$
&$(N-N_{\infty})=[c_{1}\pm\frac{3c_{o}}{2}(t-t_{o})]^\frac{2}{3}$\\
$\mathcal{F}_m\infty$ & $[\frac{\pi}{2},-\frac{\pi}{4},\frac{\pi}{2}]$
&$[\frac{1}{\sqrt{2}},0,0]$
&$(N-N_{\infty})=[c_{1}\pm\frac{3c_{o}}{2}(t-t_{o})]^\frac{2}{3}$ \\
$\mathcal{M}_m\infty$ & $[\frac{\pi}{2},\frac{\pi}{2},\phi]$ & $[0,0,\chi]$ & \\
&  &$y\rightarrow \pm\infty$
$K\rightarrow \mp\infty$& $a=a_{0}exp[\gamma(t-t_{0})]$\\
&  &$y\rightarrow +\infty$
$K\rightarrow +\infty$& $(N-N_{\infty})=[c_{1}\pm\frac{c_{o}}{2}(t-t_{o})]^{2}$\\
& &$y\rightarrow -\infty$
$K\rightarrow -\infty$& $(N-N_{\infty})=[c_{1}\pm\frac{3c_{o}}{2}(t-t_{o})]^{\frac{2}{3}}$\\
$\mathcal{O}_m\infty$ & $[{\small arccot}(\chi),-\frac{\pi}{2},\phi]$ & $[0,0,f(\phi)>0 \;\; \forall \phi ]$ &   \\
&  &$y\rightarrow \pm\infty$
$K\rightarrow \mp\infty$& $a=a_{0}exp[\gamma(t-t_{0})]$\\
&  &$y\rightarrow +\infty$
$K\rightarrow +\infty$& $(N-N_{\infty})=[c_{1}\pm\frac{c_{o}}{2}(t-t_{o})]^{2}$\\
& &$y\rightarrow -\infty$
$K\rightarrow -\infty$& $(N-N_{\infty})=[c_{1}\pm\frac{3c_{o}}{2}(t-t_{o})]^{\frac{2}{3}}$\\
\br
\end{tabular}
\end{center}
\end{table}

\begin{table}\caption{Stability of the fixed points in non-vacuum
$exp\,(-R/{\Lambda})$-gravity.}\label{tab asy matter}
\begin{center}
\begin{tabular} {lp{1.3in}ll}
\br
{\small Point} &{\small Stability}\\
\mr
$\mathcal{A}_{\infty}$ & {\small Stable} \\
$\mathcal{B}_{\infty}$ & {\small Unstable} \\
$\mathcal{C}_{\infty}$ & {\small Unstable} \\
$\mathcal{D}_{\infty}$ & {\small Unstable} \\
$\mathcal{E}_{\infty}$ & {\small Stable }\\
$\mathcal{F}_{\infty}$ & {\small Unstable }\\
\br
\end{tabular}
\end{center}
\end{table}

\section{Discussion and Conclusions}\label{conclusioni}
In this paper we have applied the dynamical system approach to the
exponential gravity cosmological model, and found exact solutions
together with their stability for both the vacuum and matter cases.

In the vacuum case we found four finite critical points $\mathcal{A}_{v}$, $\mathcal{B}_{v}$,
$\mathcal{C}_{v}$ and $\mathcal{D}_{v}$, of which only two $\mathcal{C}_{v}$ and $\mathcal{D}_{v}$ are found to be physical. These last points were found to represent a solution whose
nature depends on the parameter $\gamma(\Lambda)$; for $\Lambda>0$
we can have either exponential expansion $(\gamma>0)$ or exponential
contraction $(\gamma<0)$ and  for $\Lambda<0$ the solution oscillates.

From the stability point of view, the point $\mathcal{C}_{v}$, which lies in the
invariant submanifold $z=0$, is of particular interest because, since it is non-hyperbolic, 
it represent an attractor for $z>0$ and saddle
for $z<0$, while the other physical point $\mathcal{D}_{v}$ is found to be a
saddle.

On the other hand, the solution connected with the non-physical points $\mathcal{A}_{v}$
and $\mathcal{B}_{v}$ are found to correspond to  power law evolution and are also interesting because the orbits can approach arbitrarily close to them.

 In the asymptotic regime all the critical points represent solutions which have a maximum
 value for the scale factor, hence all models that evolve to one
 of the asymptotic future attractors will re-collapse.

The invariant submainfold $z=0$ divides the phase space into two
regions, $z>0$ and $z<0$ which correspond to $\Lambda>0$ and
$\Lambda<0$ respectively. The fact that no orbit can cross the plane
$z=0$ is then consistent with the fact that $\Lambda$ is a fixed
parameter for this model.

In the vacuum case, we found that the region $z<0$ does not contain
any finite critical point. However, in the plane $z=0$, we have the
physical point $\mathcal{C}_{v}$ which represents a de-Sitter saddle and the
non-physical points $\mathcal{A}_{v}$ and $\mathcal{B}_{v}$ 
are a repeller and saddle point respectively. Thus, the only attractors in 
the region $z<0$ are asymptotic, which means that all the models that begin their
evolution  in this region will re-collapse. It is also possible that
one can choose initial conditions in such a way that both de-Sitter
and  power law phases are present in the evolution of the model.

From a physical point of view, the region $z>0$ appears  to be more
interesting because the point $\mathcal{C}_{v}$ represents a de-Sitter 
attractor for $z>0$. It follows that there are two different possible
solutions towards which models can evolve.

Since the point $\mathcal{D}_{v}$, which represents an unstable de-Sitter phase,
lies in the region $z>0$,  a set of initial conditions exist for which orbits describe 
an intermediate de-Sitter phase (see figure \ref{z-y}). Furthermore, for models that evolve near the
non-physical point $\mathcal{B}_{v}$ an intermediate power law phase is also
present.

By looking at Figure 3 it is clear that the de-Sitter phases $\mathcal{C}_{v}$ and
$\mathcal{D}_{v}$  are separated from the past attractor $\mathcal{A}_{v}$ by the plane $q=0$,
therefore any model that starts near the past attractor $\mathcal{A}_{v}$ and
evolves toward the future de-Sitter attractor $\mathcal{C}_{v}$ will cross the
plane $q=0$, indicating a transition  from an accelerating evolution to a 
decelerating one.

The introduction of matter into this model increases the
dimensionality  of the phase space, making it more difficult to
visualize. On the bases of the relative stability of the fixed points
and the invariant submanifold structure it is possible to catalog
the possible evolutions of this model in 5 classes (see Table
\ref{model}).

\begin{table}\caption{ The sectors and the behaviour in each one for the
$exp\,(-R/\Lambda)$ gravity in matter case. Here $+/-$ corresponds to a positive/negative
 values of the coordinates.}\label{model}
\begin{center}\begin{tabular}{lllll}
\br
y & D & z & K& {\small Behaviour}  \\
\mr
+&-&-&-&$\alpha$\\
+&+&-&-&$\alpha$\\
+&-&+&-&$\alpha$\\
+&-&-&+&$\alpha$\\
+&+&+&-&$\alpha$\\
+&-&+&+&$\alpha$\\
+&+&-&+&$\alpha$\\
+&+&+&+&$\alpha$\\
\br
\end{tabular} \quad
\begin{tabular}{lllll}
\br
y & D & z & K& {\small Behaviour}  \\
\mr
-&-&-&-&$\beta$\\
-&-&-&+&$\beta$\\
-&-&+&-&$\gamma$\\
-&-&+&+&$\gamma$\\
-&+&+&-&$\epsilon$\\
-&+&+&+&$\epsilon$\\
-&+&-&-&$\delta$\\
-&+&-&+&$\delta$\\
\br
\end{tabular}
\end{center}
\end{table}

The first class ($\alpha$) is characterized  by the fact that that the
cosmic histories evolve towards re-collapse. This class also contains
cosmic histories which include an intermediate almost power law
transient phase. For the second class ($\beta$), cosmic histories evolve
towards re-collapse, but in addition to a power law transient phase there can be 
also an oscillating one. The third class ($\gamma$) contains two types of orbits. 
Depending on the initial conditions, the models will either evolve towards re-collapse 
or towards a de-Sitter type solution. During this evolution, a  transient de Sitter phase 
can be present. A fourth class ($\epsilon$) also contains two types of orbits. The 
universe can either re-collapse or evolve  to a de-Sitter type model. In this case there are transient phases that  include two different unstable de-Sitter evolutions . The final  class ($\delta$) contains models that either re-collapse or end up at an oscillating solution. The possible transient phase in this class are multiple oscillations and/or almost power law behaviour.
As it can be seen from Table \ref{model}, the condition $D/z>0$ coming from the
weak energy condition, required  for standard matter, allows one
to exclude completely the classes $\gamma$ and $\delta$.

In conclusion,  $exp\,(-\frac{R}{\Lambda})$ gravity has a very rich
structure that includes a series of diverse
cosmological histories. Particularly important are the ones including
multiple de Sitter phases because they could provide us with natural 
models describing the early and late time acceleration of the Universe.
Unfortunately, as is clear from  Figure \ref{z-y}, this
scenario does not include a decelerated expansion phase between these two
de-Sitter phases, so unless some other mechanism is taken into
account in these cosmic histories will not admit a standard structure
formation scenario.

\appendix
\section{}

In section $3.1$ we used the center manifold theorem to analyze the
stability of the non--hyperbolic fixed point $\mathcal{C}_{v}$ in
the vacuum case, here we will give a brief review of this approach
$\cite{9}$. Consider the nonlinear system (bold letters represent
vectors)
\begin{eqnarray}
\dot \textbf{u}=f(\textbf{u}).
\end{eqnarray}
For simplicity we shall assume that the origin is a non--hyperbolic
fixed point for this system (this assumption does not affect the
generality of our treatment because it is always possible to change
the coordinates to make the fixed point the origin of the new
coordinate system). If $f\in {C}^{1}(E)$ and
$f(\textbf{0})=\textbf{0}$, then this system can be written in the
diagonal form
\begin{eqnarray}
\dot \textbf{u}=J \textbf{u}+T(\textbf{u})\;,
\end{eqnarray}
where $J=Df(\textbf{0})=diag[Z,P,N]$, the square matrices $Z,P,N$
have $r$ eigenvalues of zero real part, $p$ eigenvalues of positive
real part and $n$ eigenvalues of negative real part respectively and
$T(\textbf{u})=f(\textbf{u})-J\textbf{u}$, where $T\in
\textbf{C}^{1}(E)$, $T(\textbf{0})=\textbf{\textbf{0}}$ and $D
T(\textbf{0})=\textbf{0}$. The system $(48)$ can be divided into
three subsystems
\begin{eqnarray}\nonumber
\dot \textbf{x}=Z \textbf{x}+F(\textbf{x},\textbf{y},\textbf{z}),\\
\dot \textbf{y}=P
\textbf{y}+G(\textbf{x},\textbf{y},\textbf{z}),\\\nonumber \dot
\textbf{z}=N
\textbf{z}+H(\textbf{x},\textbf{y},\textbf{z}),\nonumber
\end{eqnarray}
where $(\textbf{x},\textbf{y},\textbf{z})\in R^{r}\times R^{p}\times
R^{n}$, $\textbf{F}(\textbf{0})= \textbf{G}(\textbf{0})=
\textbf{H}(\textbf{0})=\textbf{\textbf{0}}$, and
$D\textbf{F}(\textbf{0})=
D\textbf{G}(\textbf{0})=D\textbf{H}(\textbf{0})=\textbf{0}$. If
$(F,G,H)\in C^{m}(E)$ with $m\geq 1$, it follows from the local
center manifold theory that there exist a $z$-dimensional invariant
center manifold $W^{c}_{local}(\textbf{0})$ defined by
\begin{eqnarray*}
\fl\textbf{W}^{c}_{local}(\textbf{0})=
\left\{(\textbf{x},\textbf{y},\textbf{z})\in R^{r}\times R^{p}\times
R^{n}|
\textbf{y}=\textbf{h1}(\textbf{x}),\textbf{z}=\textbf{h2}(\textbf{x})\;\textrm{for}
|x|<\delta \right\}\;,
\end{eqnarray*}
for some $\delta > 0 $, where $(\textbf{h1},\textbf{h2})\in
C^{r}(N_{\delta}(\textbf{0}))$,
$\textbf{h1}(\textbf{0})=\textbf{h2}(\textbf{0})=\textbf{0}$,
$D\textbf{h1}(\textbf{0})=D\textbf{h2}(\textbf{0})=\textbf{0}$, and
they satisfy
\begin{eqnarray*}
\fl D\textbf{h1}(\textbf{x})[Z\textbf{x}+F(\textbf{x},\textbf{h1}(\textbf{x}),\textbf{h2}(\textbf{x}))]-P\textbf{h1}(\textbf{x})-G(\textbf{x},\textbf{h1}(\textbf{x}),\textbf{h2}(\textbf{x}))=0\;,\\
\fl
D\textbf{h2}(\textbf{x})[Z\textbf{x}+F(\textbf{x},\textbf{h1}(\textbf{x}),\textbf{h2}(\textbf{x}))]
-N\textbf{h2}(\textbf{x})-H(\textbf{x},\textbf{h1}(\textbf{x}),\textbf{h2}(\textbf{x}))=0\;.
\end{eqnarray*}
In the neighborhood of a non-hyperbolic fixed point the
qualitative behavior of the system $(48)$ is equivalent to the
qualitative behavior of the reduced system
\begin{eqnarray}\label{EqW0}
\dot \textbf{x}=Z
\textbf{x}+F(\textbf{x},\textbf{h1}(\textbf{x}),\textbf{h2}(\textbf{x})).
\end{eqnarray}
The functions $\textbf{h1}(\textbf{x}), \textbf{h2}(\textbf{x})$ can
be approximated by substituting the series expansion of their
components into equations $(51)$ and $(52)$. In the case when we have a double zero eigenvalues the coordinate which correspond to the non-zero eigenvalue can be approximated by the function, 
\begin{eqnarray}
\textbf{h}(\textbf{x}_1,\textbf{x}_2)=a\,\textbf{x}_1^2+b\,\textbf{x}_1^2 \textbf{x}_2^2+c\,\textbf{x}_2^2+0(|\textbf{x}|^3),
\end{eqnarray}
where $\textbf{x}_1$ and $\textbf{x}_2$ are the coordinates which correspond to the zero eigenvalues. In general the flow on the center manifold near the fixed point takes the form
\begin{eqnarray}\label{EqW0appr}
\dot{\textbf{x}}=a\,\textbf{x}^r+...,
\end{eqnarray}
If $r\geq2$ and $a_{r}\neq0$, then for $r$ even we have saddle-node
at the fixed point, for $r$ odd and $a_{r}>0$ we have unstable node
and for $r$ odd and  $a_{r}<0$ we have a topological saddle.\newpage

\section*{Acknowledgments}

This work was supported by the National Research Foundation (South
Africa) and the Ministrero degli Affari Esteri-DG per la Promozione
e Cooperazione Culturale (Italy) under the joint Italy/South Africa
science and technology agreement. M.A thanks the African Institute for Mathematical 
Sciences (AIMS) for financial support.



\section*{References}
\begin{thebibliography}{999}
\bibitem{b}Jassa HK, Bagla JS and Padmanabhan T 2005 {\it Phys. Rev.} {\bf D72} 103503 arXiv:astro-ph/0506748

\bibitem{a} Carloni S, Dunsby PKS, Capozziello S and  Troisi  2005 {\it Class. Quantum Grav.} {\bf22} 4839 arXiv:gr-qc/0410046

\bibitem{generaldynsys} Carloni S, Troisi A, Dunsby PKS arXiv:0706.0452 

\bibitem{Starobinsky}  Starobinsky AA 1980 {\it Phys. Lett.} {\bf B91} 99

\bibitem{Capozziello} Capozziello S, Carloni S and Troisi A 2003 {\it Recent Res. Devel.Astronomy \& Astrophysics} {\bf1}
625 arXiv: astro-ph/0303041

\bibitem{c}  Kerner R 1982 \grg {\bf14} 453 ;
Duruisseau J P, Kerner R 1986 \cqg {\bf3} 817

\bibitem{d} Teyssandier P 1989 {\it Class. Quantum Grav.} {\bf6} 219

\bibitem{e} Magnano G, Ferraris M and Francaviglia M 1987 {\it Gen. Rel. Grav.} {\bf19} 465

\bibitem{f} Capozziello S, Cardone V.F, Carloni S, Troisi A 2004 {\it Phys. Lett.} A{\bf A326} 292 arXiv:gr-qc/0404114

\bibitem{g} Capozziello S, de Ritis R and Marino A A 1998 {\it Gen. Rel. Grav.} {\bf 30} 1247 arXiv:gr-qc/9804053

\bibitem{8} Capozziello S, Cardone VF and Troisi A 2006 {\it J.Cosmol. Astropart. Phys.} JCAP{\bf0608} 001 arXiv:astro-ph/0602349

\bibitem{9} Perko L 1996 {\it Differential Equations and Dynamical Systems }(New York:Springer-Verlag )

\bibitem{10} Barrow JD and Cotsakis S 1988 {\it Phys. Lett.} {\bf B214} 515

\bibitem{11} Barrow JD 1988 {\it Nucl. Phys.} {\bf B296} 697

\bibitem{12} Schmidt HJ gr-qc/0407095; M\"{u}ller V and Schmidt HJ 1985 {\it Gen. Rel. Grav.} {\bf17} 769

\bibitem{13} Clifton T and Barrow JD 2006 {\it Class. Quant. Grav.} {\bf23} 2951

\bibitem{14} Barrow JD and Clifton T 2005 {\it Class. Quant. Grav.} {\bf22} L1

\bibitem{15} Clifton T and Barrow JD 2005 {\it Phys. Rev.} {\bf D72} 103005

\bibitem{16} Barrow JD and Hervik S 2006 {\it Phys. Rev.} {\bf D73} 023007

\bibitem{17} S. Deser and B. Tekin 2003 {\it Phys. Rev.} {\bf D67} 084009

\bibitem{18} Bach R 1921{\it Math. Zeitschrift} {\bf9} 110

\bibitem{19} Coley AA and Hervik S 2004 {\it Class. Quant. Grav.} {\bf 21} 4193;
Hervik S, van den Hoogen RJ and Coley AA 2005 {\it Class. Quant. Grav.} {\bf 22} 607; Hervik S, van den Hoogen RJ, Lim WC and Coley AA 2006 {\it Class. Quant. Grav.} {\bf 23} 845; Hervik S and  Lim WC  2006 {\it Class. Quant. Grav.} {\bf 23} 3017.

\bibitem{20} Leach JA Carloni S and Dunsby PKS 2006 {\it Class. Quant. Grav.} {\bf23} 4915 arXiv: gr-qc/0603012

\bibitem{21}Cotsakis S, Demaret J, De Rop Y and Querella L 1993 {\it Phys.
Rev.} D{\bf48} 4595; Demaret J and Querella L 1995 {\it Class. Quantum Grav.} {\bf12} 3085

\bibitem{25} Starobinskii AA 1983 {\it Sov. Phys. JETP Lett.} {\bf 37} 66

\bibitem{26} Jensen LG and Stein-Schabes J 1987 {\it Phys. Rev.} {\bf D35} 1146

\bibitem{27} Wald R 1983 {\it Phys. Rev.} {\bf D28} 2118

\bibitem{28} Barrow JD 1987 {\it Phys. Lett.} {\bf B187} 12

\bibitem{29} Dabrowski MP 2006 {\it Annalen Phys.} {\bf15} 352 arXiv:astro-ph/0606574

\bibitem{30} Barrow JD 1978 {\it Nature.} {\bf272} 211

\bibitem{31} Barrow JD 1977 {\it Mon. Not. R. astr. Soc.} {\bf178} 625;
Collins CB 1974 {\it Comm. Math. Phys.} {\bf39} 131; Ellis GFR and Collins CB
1979 {\it Phys. Rep.} {\bf56} 65; Shikin IS 1976 {\it Sov. Phys.} JETP{\bf41} 794

\bibitem{32} Barrow JD 1997 {\it Phys. Rev.} {\bf D55} 7451

\bibitem{3} Barrow JD and Ferreira P 1997 {\it Phys. Rev. Lett.} {\bf78} 610

\bibitem{34} Kaloper N 1991 {\it Phys. Rev.} {\bf D44} 2380

\bibitem{35}Barrow JD 2004 {\it Class. Quantum Grav.} {\bf21} L79 arXiv:gr-qc/0403084

\bibitem{36} Barrow JD 2004 {\it Class. Quantum Grav.} {\bf21} 5619

\bibitem{37} Barrow JD and Tsagas CG 2005 {\it Class. Quantum Grav.} {\bf22} 1563

\bibitem{38}  Cotsakis S and Klaoudatou I 2005 {\it J. Geom. Phys.} {\bf55} 306 arXiv:gr-qc/0409022

\bibitem{39} Gr{\o }n \O  and Hervik S 2003 {\it Int. J. Theo. Ph. Gr. Th. Non-L.
Opt.} {\bf10} 29 arXiv:gr-qc/0205026

\bibitem{40} Barrow JD and Hervik S 2002 {\it Class.Quant.Grav.} {\bf19}, 5173

\bibitem{41} Belinskii V Lifshitz EM and Khalatnikov I 1970 {\it Adv. Phys.} {\bf19} 525

\bibitem{42} Chernoff D and Barrow JD 1983 {\it Phys. Rev. Lett.} {\bf 50} 134

\end{thebibliography}
\end{document}